\documentclass[12pt,  draftclsnofoot, journal, onecolumn]{IEEEtran}
\usepackage{amsmath,amsfonts}
\usepackage{algorithmic}
\usepackage{algorithm}
\usepackage{array}
\usepackage[caption=false,font=normalsize,labelfont=sf,textfont=sf]{subfig}
\usepackage{textcomp}
\usepackage{stfloats}
\usepackage{url}
\usepackage{verbatim}
\usepackage{graphicx}
\usepackage{cite}
\usepackage{amsmath,graphicx,url,times, amssymb, latexsym}
\begin{document}

\def\beq{\begin{equation}} 
\def\eeq{\end{equation}}
\def\beqa{\begin{eqnarray}} 
\def\eeqa{\end{eqnarray}}

\def\bear{\begin{array}} 
\def\eear{\end{array}}
\def\ben{\begin{enumerate}}
\def\een{\end{enumerate}} 
\def\bit{\begin{itemize}}
\def\eit{\end{itemize}} 
\def\bfi{\begin{figure}}
\def\efi{\end{figure}} 
\def\ltp{\left(} 
\def\rtp{\right)}
\def\ltc{\left\{} 
\def\rtc{\right\}}
\def\lts{\left[} 
\def\rts{\right]}
\def\ltb{\left|} 
\def\rtb{\right|} 

\def\teq{\triangleq}

\def\dsp{\displaystyle}
\def\bel{\label}

\makeatletter
\def\BState{\State\hskip-\ALG@thistlm}
\makeatother

\def\erf#1{(\ref{#1})}
\def\merf#1#2{(\ref{#1}--\ref{#2})}
\def\diag{\mbox{$\text{diag}$}}
\def\real{\mbox{$\text{Re}$}}
\def\imag{\mbox{$\text{Im}$}}
\def\bAt{\mbox{$\bA^T$}}
\def\bDk{\mbox{$\bSigma_{m}$}}
\def\bDuk{\mbox{$\underline{\bSigma}_{m}$}}
\def\bmutk{\mbox{$\tilde{\bmu}_m$}}
\def\Ast{\mbox{$\bA^{(s)T}$}}
\def\skn{\mbox{$\sigma_{kn}^2$}}
\def\stkns{\mbox{$\tilde{\sigma}_{kn}^{(s)\,2}$}}
\def\mutknssq{\mbox{$\tilde{\mu}_{kn}^{(s)\,2}$}}
\def\muhk{\mbox{$\hat{\mu}_k$}}
\def\bAr{\mbox{$\bA_r$}}
\def\bArt{\mbox{$\bA_r^T$}}
\def\rth{\mbox{$r^{\text{th}}$}}
\def\Arst{\mbox{$\bA_r^{(s)\,T}$}}
\def\tth{\mbox{$t^{\text{th}}$}}
\def\xhts{\mbox{$\hat{x}_t^{(s)}$}}
\def\bal{\mbox{$\mathbf{\alpha}$}}
\def\Ls{\mbox{$\mathcal{L}$}}
\def\Lk{\mbox{$\Lambda_k$}}
\def\Lst{\mbox{$\Ls_t$}}
\def\Lstk{\mbox{$\Ls_{k,t}$}}
\def\Pskt{\mbox{$\Ps_{k,t}$}}

\def\xtn{\mbox{$x_{tn}$}}

\def\Ps{\mbox{$\mathcal{P}$}}
\def\Pst{\mbox{$\Ps_t$}}

\def\xhtns{\mbox{$\hat{x}_{tn}^{(s)}$}}
\def\xht{\mbox{$\hat{x}_t$}}

\def\gij{\mbox{$g_{ij}$}}
\def\anm{\mbox{$a_{nm}$}}
\def\adj{\mbox{$\text{adj}\,$}}
\def\danm{\mbox{$\delta_\alpha[n,m]$}}
\def\hapnm{\mbox{$h_{\alpha\rho}[n,m]$}}
\def\cA{\mbox{$\mathcal{A}$}}
\def\cB{\mbox{$\mathcal{B}$}}
\def\cC{\mbox{$\mathcal{C}$}}
\def\cD{\mbox{$\mathcal{D}$}}
\def\cF{\mbox{$\mathcal{F}$}}
\def\cH{\mbox{$\mathcal{H}$}}
\def\cL{\mbox{$\mathcal{L}$}}
\def\cT{\mbox{$\mathcal{T}$}}
\def\calX{\mbox{$\mathcal{X}$}}
\def\calY{\mbox{$\mathcal{Y}$}}
\def\calZ{\mbox{$\mathcal{Z}$}}
\def\calm{\mathcal{M}}

\def\Az{\mbox{$A(z)$}}
\def\Bz{\mbox{$B(z)$}}
\def\Gz{\mbox{$G(z)$}}
\def\rhob{\mbox{$\rho_{b}$}}
\def\thetab{\mbox{$\theta_{b}$}}
\def\Real{\mbox{$\text{Re}\,$}}
\def\Imag{\mbox{$\text{Im}\,$}}

\def\bT{\mbox{$\mathbf{T}$}}
\def\bD{\mbox{$\mathbf{D}$}}
\def\bG{\mbox{$\mathbf{G}$}}
\def\bbG{\mbox{$\mathsf{G}$}}
\def\bbGrls{\mbox{$\bbG_{\text{RLS}}$}}
\def\bGrls{\mbox{$\bG_{\text{RLS}}$}}
\def\bbGk{\mbox{$\bbG_{\text{K}}$}}
\def\bmu{\mbox{$\boldsymbol{\mu}$}}
\def\bvarphi{\mbox{$\boldsymbol{\varphi}$}}
\def\tauh{\mbox{$\hat{\tau}$}}
\def\btheta{\mbox{$\boldsymbol{\theta}$}}
\def\bdelta{\mbox{$\boldsymbol{\delta}$}}
\def\bSigma{\boldsymbol{\Sigma}}
\def\bd{\mbox{$\mathbf{d}$}}
\def\taub{\mbox{$\bar{\tau}$}}
\def\epsb{\mbox{$\bar{\epsilon}$}}
\def\btau{\mbox{$\boldsymbol{\tau}$}}
\def\btaub{\mbox{$\bar{\boldsymbol{\tau}}$}}
\def\btauh{\mbox{$\hat{\boldsymbol{\tau}}$}}
\def\bQ{\mbox{$\mathbf{Q}$}}
\def\bbQ{\mbox{$\mathsf{Q}$}}
\def\bbD{\mbox{$\mathsf{D}$}}
\def\bg{\mbox{$\mathbf{g}$}}
\def\blambda{\mbox{$\mathbf{\lambda}$}}
\def\cO{\mbox{$\mathcal{O}$}}
\def\cM{\mbox{$\mathcal{M}$}}
\def\bF{\mbox{$\mathbf{F}$}}
\def\bbF{\mbox{$\mathsf{F}$}}
\def\bZ{\mbox{$\mathbf{Z}$}}
\def\bbZ{\mbox{$\mathsf{Z}$}}
\def\bZfb{\mbox{$\bZ_{\text{fb}}$}}
\def\bz{\mbox{$\mathbf{z}$}}
\def\bb{\mbox{$\mathbf{b}$}}
\def\bbf{\mbox{$\mathbf{f}$}}
\def\bB{\mbox{$\mathbf{B}$}}
\def\bbB{\mbox{$\mathsf{B}$}}
\def\bM{\mbox{$\mathbf{M}$}}
\def\bbm{\mbox{$\mathbf{m}$}}
\def\br{\mbox{$\mathbf{r}$}}
\def\bxh{\mbox{$\hat{\bx}$}}
\def\bX{\mbox{$\mathbf{X}$}}
\def\bbX{\mbox{$\mathsf{X}$}}
\def\bR{\mbox{$\mathbf{R}$}}
\def\bA{\mbox{$\mathbf{A}$}}
\def\bbR{\mbox{$\mathsf{R}$}}
\def\bRx{\mbox{$\bR_{\mathbf{x}}$}}
\def\bY{\mathbf{Y}}
\def\bPhi{\mbox{$\mathbf{\Phi}$}}
\def\bzeta{\mbox{$\boldsymbol{\zeta}$}}
\def\bbeta{\mbox{$\boldsymbol{\eta}$}}
\def\etab{\mbox{$\boldsymbol{\eta}$}}
\def\bP{\mbox{$\mathbf{P}$}}
\def\bbP{\mbox{$\mathsf{P}$}}
\def\bPz{\mbox{$\bP_{\mathbf{Z}}$}}
\def\bPc{\mbox{$\bP_{\mathbf{C}}$}}
\def\bPcperp{\mbox{$\bP_{\mathbf{C}}^\perp$}}
\def\bV{\mbox{$\mathbf{V}$}}
\def\bx{\mbox{$\mathbf{x}$}}
\def\by{\mbox{$\mathbf{y}$}}
\def\be{\mbox{$\mathbf{e}$}}
\def\bLambda{\mbox{$\mathbf{\Lambda}$}}
\def\bN{\mbox{$\mathbf{N}$}}
\def\bv{\mbox{$\mathbf{v}$}}
\def\bvk{\mbox{$\mathbf{v}_{\mathbf{k}}$}}
\def\bk{\mbox{$\mathbf{k}$}}
\def\bK{\mbox{$\mathbf{K}$}}
\def\bbK{\mbox{$\mathsf{K}$}}
\def\bS{\mbox{$\mathbf{S}$}}
\def\brhon{\mbox{$\brho_{\mathbf{n}}$}}
\def\brhoni{\mbox{$\brho_{\mathbf{n}}^{-1}$}}
\def\bSQ{\mbox{$\bS_{\mathbf{Q}}$}}
\def\bSc{\mbox{$\bS_{\mathbf{c}}$}}
\def\bSx{\mbox{$\bS_{\mathbf{x}}$}}
\def\bgz{\mbox{$\bg_{\mathbf{Z}}$}}
\def\bSz{\mbox{$\bS_{\mathbf{z}}$}}
\def\bSn{\mbox{$\bS_{\mathbf{n}}$}}
\def\bSf{\mbox{$\bS_{\mathbf{f}}$}}
\def\bSs{\mbox{$\bS_{\mathbf{s}}$}}
\def\bI{\mbox{$\mathbf{I}$}}
\def\bw{\mbox{$\mathbf{w}$}}
\def\bff{\mbox{$\mathbf{f}$}}
\def\bW{\mbox{$\mathbf{W}$}}
\def\bbW{\mbox{$\mathsf{W}$}}
\def\bU{\mbox{$\mathbf{U}$}}
\def\bbU{\mbox{$\mathsf{U}$}}
\def\bu{\mbox{$\mathbf{u}$}}
\def\bp{\mbox{$\mathbf{p}$}}
\def\bpz{\mbox{$\mathbf{p}_{\mathbf{z}}$}}
\def\bpHz{\mbox{$\mathbf{p}^T_{\mathbf{z}}$}}
\def\bc{\mbox{$\mathbf{c}$}}
\def\bss{\mbox{$\mathbf{s}$}}
\def\ba{\mbox{$\mathbf{a}$}}
\def\bH{\mbox{$\mathbf{H}$}}
\def\bh{\mbox{$\mathbf{h}$}}
\def\mth{\mbox{$m$th}}
\def\sinc{\mbox{$\text{sinc}$}}
\def\bphi{\mbox{$\mathbf{\Phi}$}}
\def\brho{\mbox{$\mathbf{\rho}$}}
\def\bzero{\mbox{$\mathbf{0}$}}
\def\mini{\mbox{$\text{minimize}$}}
\def\subj{\mbox{$\text{subject to}$}}
\def\st{\mbox{$\text{s.t.}$}}
\def\nth{\mbox{$n$th}}
\def\kth{\mbox{$k$th}}
\def\lth{\mbox{$l$th}}
\def\dth{\mbox{$d$th}}
\def\ith{\mbox{$i$th}}
\def\jth{\mbox{$j$th}}
\def\Kth{\mbox{$K$th}}
\def\bxtn{\mbox{$\tilde{\bx}_n$}}
\def\bwtn{\mbox{$\tilde{\bw}_n$}}
\def\bWtn{\mbox{$\tilde{\bW}_n$}}
\def\bXtn{\mbox{$\tilde{\bX}_n$}}
\def\bXt{\mbox{$\tilde{\bX}$}}
\def\bl{\mbox{$\mathbf{l}$}}
\def\bL{\mbox{$\mathbf{L}$}}
\def\bXtfb{\mbox{$\bXt_{\text{fb}}$}}
\def\bXtHfb{\mbox{$\bXt^T_{\text{fb}}$}}
\def\bYt{\mbox{$\tilde{\bY}$}}
\def\byt{\mbox{$\tilde{\by}$}}
\def\bFm{\mbox{$\bF_M$}}
\def\bFmi{\mbox{$\bF_M^{-1}$}}
\def\bFmh{\mbox{$\bF_M^T$}}
\def\bwdq{\mbox{$\bar{\bw}_{\text{dq}}$}}
\def\bwdqH{\mbox{$\bar{\bw}_{\text{dq}}^T$}}
\def\balpha{\mbox{$\boldsymbol{\alpha}$}}
\def\bepsilon{\mbox{$\boldsymbol{\epsilon}$}}

\def\tr{\mbox{$\text{tr}$}}
\def\bJ{\mbox{$\mathbf{J}$}}
\def\xtmn{\mbox{$\tilde{x}_m[n]$}}

\def\bOmega{\mbox{$\mathbf{\Omega}$}}

\def\bwnn{\mbox{$\text{BW}_{\text{NN}}$}}
\def\rnn{\mbox{$r_{\text{NN}}$}}

\def\bSxi{\mbox{$\bS_{\mathbf{x}}^{-1}$}}
\def\bSzi{\mbox{$\bS_{\mathbf{z}}^{-1}$}}
\def\bSni{\mbox{$\bS_{\mathbf{n}}^{-1}$}}
\def\bSxt{\mbox{$\bS_{\mathbf{x}}^T$}}
\def\bSxh{\mbox{$\hat{\bS}_{\mathbf{x}}$}}
\def\bC{\mbox{$\mathbf{C}$}}
\def\bxi{\mbox{$\boldsymbol{\xi}$}}
\def\bnu{\mbox{$\boldsymbol{\nu}$}}
\def\bbC{\mbox{$\mathsf{C}$}}
\def\bbCk{\mbox{$\mathsf{C}_{\text{K}}$}}
\def\bbCkH{\mbox{$\mathsf{C}^T_{\text{K}}$}}
\def\bbCrls{\mbox{$\bbC_{\text{RLS}}$}}
\def\bbCrlsH{\mbox{$\bbC^T_{\text{RLS}}$}}
\def\bone{\mbox{$\mathbf{1}$}}
\def\bCx{\mbox{$\bC_{\mathbf{x}}$}}
\def\bCxi{\mbox{$\bC_{\mathbf{x}}^{-1}$}}
\def\bCxt{\mbox{$\bC_{\mathbf{x}}^T$}}
\def\hideal{\mbox{$h_{\text{ideal}}$}}
\def\bSh{\mbox{$\hat{\bS}$}}
\def\bSt{\mbox{$\tilde{\bS}$}}
\def\bStx{\mbox{$\tilde{\bS}_{\mathbf{x}}$}}
\def\bStxi{\mbox{$\tilde{\bS}_{\mathbf{x}}^{-1}$}}
\def\bShn{\mbox{$\hat{\bS}_{\mathbf{n}}$}}
\def\bShx{\mbox{$\hat{\bS}_{\mathbf{x}}$}}
\def\bShz{\mbox{$\hat{\bS}_{\mathbf{z}}$}}
\def\bShni{\mbox{$\hat{\bS}^{-1}_{\mathbf{n}}$}}
\def\bShxi{\mbox{$\hat{\bS}^{-1}_{\mathbf{x}}$}}
\def\bShzi{\mbox{$\hat{\bS}^{-1}_{\mathbf{z}}$}}
\def\bShxfb{\mbox{$\hat{\bS}_{\mathbf{x},\text{fb}}$}}
\def\bShnfb{\mbox{$\hat{\bS}_{\mathbf{n},\text{fb}}$}}
\def\nn{\nonumber}
\def\bl{\mathbf{l}}

\def\qk{\mbox{$w_m$}}
\def\qik{\mbox{$w_{k|i}$}}
\def\zkt{\mbox{$x_{k,t}$}}
\def\zkts{\mbox{$x_{k,t}^{(s)}$}}
\def\zmk{\mbox{$x_{m,k}$}}
\def\zmks{\mbox{$x_{m,k}^{(s)}$}}

\def\cm{\mbox{$c_{m}$}}
\def\cmk{\mbox{$c_{m,k}$}}
\def\cmks{\mbox{$c_{m,k}^{(s)}$}}

\def\Lamk{\mbox{$\Lambda_k$}}
\def\Lami{\mbox{$\Lambda_i$}}
\def\Lamik{\mbox{$\Lambda_{ik}$}}
\def\oneHalf{\mbox{$\frac{1}{2}$}}
\def\oneFourth{\mbox{$\frac{1}{4}$}}
\def\oneSixth{\mbox{$\frac{1}{6}$}}
\def\mukt{\mbox{$\mu_{k}^{T}$}}
\def\mukrstar{\mbox{$\mu^{*}_{kr}$}}
\def\mutkst{\mbox{$\tilde{\mu}_{k}^{(s)T}$}}
\def\muk{\mbox{$\mu_k$}}
\def\Qk{\mbox{$Q_k^{-1}$}}
\def\xt{\mbox{$x_t$}}
\def\X{\mbox{$\mathcal{X}$}}
\def\Y{\mbox{$\mathcal{Y}$}}
\def\Ys{\mbox{$\mathcal{Y}^{(s)}$}}
\def\cY{\mbox{$\mathcal{Y}$}}
\def\cW{\mbox{$\mathcal{W}$}}
\def\cP{\mbox{$\mathcal{P}$}}
\def\cX{\mbox{$\mathcal{X}$}}
\def\cN{\mbox{$\mathcal{N}$}}
\def\cE{\mbox{$\mathcal{E}$}}
\def\cJ{\mbox{$\mathcal{J}$}}
\def\cI{\mbox{$\mathcal{I}$}}
\def\cZ{\mbox{$\mathcal{Z}$}}
\def\Z{\mbox{$\mathcal{Z}$}}
\def\ck{\mbox{$c_k$}}
\def\cks{\mbox{$c_m^{(s)}$}}
\def\ckts{\mbox{$c_{m,k}^{(s)}$}}
\def\ckt{\mbox{$c_{k,t}$}}
\def\N{\mbox{$\mathcal{N}$}}
\def\Cm{\mbox{$C_m$}}

\def\muik{\mbox{$\mu_{ik}$}}
\def\mukn{\mbox{$\mu_{kn}$}}
\def\mukl{\mbox{$\mu_{kl}$}}
\def\muhks{\mbox{$\hat{\bmu}_m^{(s)}$}}
\def\muhkns{\mbox{$\hat{\mu}_{mn}^{(s)}$}}
\def\muhkn{\mbox{$\hat{\mu}_{kn}$}}
\def\mut{\mbox{$\tilde{\mu}$}}
\def\mutks{\mbox{$\tilde{\mu}_k^{(s)}$}}
\def\bmutms{\mbox{$\tilde{\boldsymbol{\mu}}_m^{(s)}$}}
\def\bmutm{\mbox{$\tilde{\mathbb{\mu}}_m^{(s)}$}}
\def\muh{\mbox{$\hat{\bmu}$}}
\def\mub{\mbox{$\breve{\bmu}$}}

\def\btwo{\mbox{$\tilde{\bw}_o$}}
\def\btwHo{\mbox{$\tilde{\bw}^T_o$}}
\def\btwa{\mbox{$\tilde{\bw}_a$}}
\def\btwHa{\mbox{$\tilde{\bw}^T_a$}}
\def\bhwa{\mbox{$\hat{\bw}_a$}}
\def\bhwHa{\mbox{$\hat{\bw}^T_a$}}

\def\bwtgsc{\mbox{$\tilde{\bw}_{\text{gsc}}$}}
\def\bwtgscH{\mbox{$\tilde{\bw}_{\text{gsc}}$}}

\def\btk{\mbox{\tt btk}}
\def\numpy{\mbox{\tt NumPy}}
\def\pygsl{\mbox{\tt PyGSL}}
\def\python{\mbox{\tt Python}}

\def\SigX{\mbox{$\Sigma_{\mathbf{X}}$}}
\def\SigY{\mbox{$\Sigma_{\mathbf{Y}}$}}
\def\SigXi{\mbox{$\Sigma^{-1}_{\mathbf{X}}$}}
\def\SigYi{\mbox{$\Sigma^{-1}_{\mathbf{Y}}$}}

\def\xls{\mbox{$\by_{1:K}^{(s)}$}}
\def\xlsT{\mbox{$\by_{1:K}^{(s)T}$}}
\def\nls{\mbox{$g_{1:K}^{(s)}$}}
\def\nts{\mbox{$g_k^{(s)}$}}
\def\wns{\mbox{$w_{1:K_{\text{w}}}^{(s)}$}}
\def\wn{\mbox{$w_{1:K_{\text{w}}}$}}
\def\cs{\mbox{$c_{1:K}^{(s)}$}}
\def\oks{\mbox{$\bo_m^{(s)}$}}
\def\xts{\mbox{$\by_k^{(s)}$}}
\def\xtsT{\mbox{$\bx_k^{(s)T}$}}
\def\sks{\mbox{$\bbs_m^{(s)}$}}
\def\skns{\mbox{$s_{kn}^{(s)}$}}
\def\dks{\mbox{$d_m^{(s)}$}}
\def\muok{\mbox{$\bmu^0_m$}}
\def\mutkns{\mbox{$\hat{\mu}_{mn}^{0(s)}$}}

\def\Dk{\mbox{$\bSigma_{m}^{-1}$}}
\def\Dik{\mbox{$\bSigma_{im}^{-1}$}}

\def\bs{\mbox{$\mathbf{s}$}}

\def\bbA{\mbox{$\underline{\bA}$}}
\def\bbSigma{\mbox{$\underline{\bSigma}$}}
\def\bbbb{\mbox{$\mathbb{b}$}}
\def\bbmutm{\mbox{$\underline{\tilde{\bmu}}_m$}}
\def\As{\mbox{$\bA^{(s)}$}}
\def\bAs{\mbox{$\bA^{(s)}$}}
\def\Ars{\mbox{$\bA_r^{(s)}$}}
\def\Aos{\mbox{$\bA_0^{(s)}$}}
\def\AsT{\mbox{$\bA^{(s)\,T}$}}
\def\ArsT{\mbox{$\bA_r^{(s)\,T}$}}
\def\cSOne{\mbox{$S^{(1)}$}}
\def\cSTwo{\mbox{$S^{(2)}$}}
\def\brs{\mbox{$b_r^{(s)}$}}
\def\bos{\mbox{$b_0^{(s)}$}}
\def\Dki{\mbox{$D_k^{-1}$}}
\def\muhko{\mbox{$\hat{\mu}_k^0$}}
\def\miT{\mbox{$m_i^T$}}
\def\mjT{\mbox{$m_j^T$}}
\def\fiT{\mbox{$f_i^T$}}
\def\ots{\mbox{$\bo_{m,k}^{(s)}$}}
\def\otsT{\mbox{$\bo_{m,k}^{(s)T}$}}
\def\oksT{\mbox{$\bo_m^{(s)T}$}}

\def\cSkrOne{\mbox{$S_{kr}^{(1)}$}}
\def\cSkTwo{\mbox{$S_k^{(2)}$}}
\def\cSkrTwo{\mbox{$S_{kr}^{(2)}$}}

\def\bsT{\mbox{$b^{(s)T}$}}
\def\brsT{\mbox{$b_r^{(s)T}$}}

\def\Dks{\mbox{$\Delta_k^{(s)}$}}
\def\DksT{\mbox{$\Delta_k^{(s)T}$}}

\def\Xs{\mbox{$\X^{(s)}$}}
\def\Zs{\mbox{$\Z^{(s)}$}}
\def\xhap{\mbox{$\hat{x}_{\alpha\rho}[n]$}}
\def\qapmn{\mbox{$q_{\alpha\rho}^{(m)}[n]$}}
\def\qapm{\mbox{$q_{\alpha\rho}^{(m)}$}}
\def\thetap{\mbox{$\theta_1$}}

\def\G{\mbox{$\mathcal{G}$}}
\def\Gn{\mbox{$\G_n$}}
\def\A{\mbox{$A$}}
\def\Aks{\mbox{$A^{(s)}_{k}$}}
\def\Akst{\mbox{$A^{(s)T}_{k}$}}
\def\Bks{\mbox{$B^{(s)}_{k}$}}
\def\Bkst{\mbox{$B^{(s)T}_{k}$}}
\def\Ak{\mbox{$A_{k}$}}
\def\Atk{\mbox{$A^{T}_{k}$}}
\def\xtst{\mbox{$x^{(s)T}_t$}}

\def\threeHalves{\mbox{$\frac{3}{2}$}}
\def\oneQuarter{\mbox{$\frac{1}{4}$}}

\long\def\gobbleup#1{}

\def\argmin{\operatornamewithlimits{argmin}}
\def\argmin{\mathop{\rm argmin}}

\def\argmax{\operatornamewithlimits{argmax}}
\def\argmax{\mathop{\rm argmax}}

\def\phih{\mbox{$\phi_{\mathbf{h}}$}}

\def\bE{\mbox{$\mathbf{E}$}}
\def\bq{\mbox{$\mathbf{q}$}}
\def\bt{\mbox{$\mathbf{t}$}}
\def\bbs{\mbox{$\mathbf{s}$}}
\def\bbbs{\mbox{$\mathbf{b}^{(s)}$}}
\def\bn{\mbox{$\mathbf{n}$}}
\def\bgs{\mathbf{g}_{1:K}^{(s)}}

\def\removal{\mbox{$\text{$\epsilon$--removal}$}}
\def\push{\mbox{$\text{push}$}}

\def\bO{\mbox{$\mathbf{O}$}}
\def\cK{\mbox{$\mathbb{K}$}}
\def\cR{\mbox{$\mathbb{R}$}}

\def\cn{\mbox{$c[n]$}}
\def\ctn{\mbox{$\tilde{c}[n]$}}
\def\chn{\mbox{$\hat{c}[n]$}}
\def\xhn{\mbox{$\hat{x}[n]$}}
\def\xh{\mbox{$\hat{x}$}}
\def\Cz{\mbox{$C(z)$}}
\def\Ctz{\mbox{$\tilde{C}(z)$}}
\def\Chz{\mbox{$\hat{C}(z)$}}
\def\Ch{\mbox{$\hat{C}$}}
\def\ahi{\mbox{$\hat{a}_i$}}
\def\bhi{\mbox{$\hat{b}_i$}}
\def\cchi{\mbox{$\hat{c}_i$}}
\def\dhi{\mbox{$\hat{d}_i$}}
\def\ai{\mbox{$a_i$}}
\def\bj{\mbox{$b_j$}}
\def\bji{\mbox{$b^{-1}_i$}}
\def\bi{\mbox{$b_i$}}
\def\ci{\mbox{$c_i$}}
\def\di{\mbox{$d_i$}}
\def\ati{\mbox{$\tilde{a}_i$}}
\def\btj{\mbox{$\tilde{b}_j$}}
\def\bhj{\mbox{$\hat{b}_j$}}
\def\bhji{\mbox{$\hat{b}^{-1}_i$}}
\def\Qz{\mbox{$Q(z)$}}
\def\lra{\mbox{$\leftrightarrow$}}
\def\calL{\mbox{$\mathcal{L}$}}
\def\K{\mbox{$\mathcal{K}$}}
\def\calE{\mbox{$\mathcal{E}$}}
\def\xn{\mbox{$x[n]$}}
\def\xmin{\mbox{$x_{\text{min}}[n]$}}
\def\calG{\mbox{$\mathcal{G}$}}

\def\Qmz{\mbox{$Q^m(z)$}}
\def\Qm{\mbox{$Q^m$}}
\def\qmn{\mbox{$q^{(m)}[n]$}}
\def\qm{\mbox{$q^{(m)}$}}
\def\sumk{\mbox{$\disp\sum_k$}}
\def\suml{\mbox{$\disp\sum_{l=0}^{L-1}$}}
\def\Fz{\mbox{$F(z)$}}
\def\Fmz{\mbox{$F^m(z)$}}
\def\Fm{\mbox{$F^m$}}
\def\ointC{\mbox{$\disp\oint_{C}$}}
\def\Sonez{\mbox{$S_1(z)$}}
\def\sonen{\mbox{$s_1[n]$}}
\def\popt{\mbox{$\hat{p}$}}

\def\betas{\mbox{$\beta^*$}}
\def\zo{\mbox{$z_0$}}
\def\zhn{\mbox{$\hat{z}_n$}}
\def\zho{\mbox{$\hat{z}_0$}}
\def\Fhz{\mbox{$\hat{F}(z)$}}
\def\Fh{\mbox{$\hat{F}$}}
\def\Fhpz{\mbox{$\hat{F}'(z)$}}
\def\Fhp{\mbox{$\hat{F}'$}}
\def\gammas{\mbox{$\gamma^*$}}

\def\complex{\mathbb{C}}
\def\compstar{\mbox{$\complex^*$}}

\def\Hz{\mbox{$H(z)$}}
\def\Xz{\mbox{$X(z)$}}
\def\Xhz{\mbox{$\hat{X}(z)$}}
\def\Xh{\mbox{$\hat{X}$}}
\def\Htz{\mbox{$\tilde{H}(z)$}}
\def\Hhz{\mbox{$\hat{H}(z)$}}
\def\Hh{\mbox{$\hat{H}$}}
\def\Ghz{\mbox{$\hat{G}(z)$}}
\def\Gh{\mbox{$\hat{G}$}}
\def\fm{\mbox{$f^{(m)}$}}

\def\bwai{\mbox{$\bw_{\text{a},i}$}}
\def\bwcai{\mbox{$\bw^*_{\text{a},i}$}}
\def\bwaone{\mbox{$\bw_{\text{a},1}$}}
\def\bwatwo{\mbox{$\bw_{\text{a},2}$}}
\def\bwcaone{\mbox{$\bw^*_{\text{a},1}$}}
\def\bwcatwo{\mbox{$\bw^*_{\text{a},2}$}}
\def\bwqi{\mbox{$\bw_{\text{q},i}$}}
\def\bwqone{\mbox{$\bw_{\text{q},1}$}}
\def\bwqtwo{\mbox{$\bw_{\text{q},2}$}}

\def\fs{\mbox{$f_{\text{s}}$}}

\def\bTheta{\mbox{$\boldsymbol{\Theta}$}}
\def \bGamma{\mbox{$\boldsymbol{\Gamma}$}}
\def \bPsi{\mbox{$\boldsymbol{\Psi}$}}

\def\bwa{\mbox{$\bw_{\text{a}}$}}

\def\lambdas{\mbox{$\lambda_{\text{s}}$}}
\def\fs{\mbox{$f_{\text{s}}$}}
\def\bks{\mbox{$\bk_{\text{s}}$}}
\def\sigw{\mbox{$\sigma_{\text{w}}^2$}}
\def\sigI{\mbox{$\sigma^2_{\text{I}}$}}
\def\rhosone{\mbox{$\rho_{\text{s}1}$}}
\def\bBc{\mbox{$\bB_{\text{c}}$}}
\def\Bc{\mbox{$B_{\text{c}}$}}
\def\Bctwo{\mbox{$B_{\text{c}}^2$}}
\def\sigmah{\mbox{$\hat{\sigma}$}}

\def\Xc{\mbox{$X_{\text{c}}$}}
\def\Xs{\mbox{$X_{\text{s}}$}}
\def\Xctwo{\mbox{$X_{\text{c}}^2$}}
\def\Xstwo{\mbox{$X_{\text{s}}^2$}}
\def\sigmactwo{\mbox{$\sigma_{\text{c}}^2$}}
\def\sigmastwo{\mbox{$\sigma_{\text{s}}^2$}}

\def\BJ{\mbox{$B_\text{J}$}}
\def\BJtwo{\mbox{$B_\text{J}^2$}}
\def\sigmaJ{\mbox{$\sigma_\text{J}$}}
\def\sigmahJ{\mbox{$\hat{\sigma}_\text{J}$}}
\def\sigmahJtwo{\mbox{$\hat{\sigma}_\text{J}^2$}}
\def\sigmaJtwo{\mbox{$\sigma_\text{J}^2$}}
\def\fJ{f_{\text{J}}}
\def\bSigmaX{\bSigma_{\mathbf{X}}}

\def\expect{\cE}

\def\x{{\mathbf x}}
\def\L{{\cal L}}

\title{Sound Field Synthesis with Acoustic Waves \footnote{Author is affiliated with Amazon Inc. USA.} \footnote{This manuscript is an expanded version of a conference paper of the same title}}

\author{Mohamed F. Mansour}




\maketitle

%
%







\begin{abstract}
We propose a  practical framework to synthesize  the broadband sound-field on a small rigid surface based on the physics of sound propagation. The sound-field is generated as a composite map of two components: the room component and the device component, with acoustic plane waves as the core tool for the generation. This decoupling of room and device components significantly reduces the problem complexity and provides accurate rendering of the sound-field.  
 We describe in detail the theoretical foundations, and efficient procedures of the implementation. 
The effectiveness of the proposed framework is established through rigorous validation under different environment setups.
\end{abstract}





\section{\label{sec:introduction}Introduction}
ound-field synthesis at a microphone array in a room is the process of synthesizing audio  at each microphone of the array from a source  signal emanating from a sound source elsewhere in the room. It  is a key task in evaluating  performance metrics of speech/audio communication  devices, as it is a cost-effective methodology for data generation to replace real data collection, which is usually a slow, expensive, and error-prone procedure.  
Acoustic modeling techniques are usually utilized to generate synthetic data to either replace or augment real data collection at a fraction of the cost. These techniques usually aim at estimating the Room Impulse Response (RIR) between two points in the room. The RIR is  either computed empirically using direct measurement, or simulated using a model for room acoustics.
Empirical methods are  in general accurate, but they are relatively expensive because of the required human labor. 

Simulation methods provide a cost effective alternative as they utilize computational acoustics rather than physical measurements. A brute-force simulation  would solve the inhomogeneous acoustic wave equation with proper boundary conditions of the room  and device surface \cite{RoomAcousticsBook}. Though theoretically viable, it  requires significant effort to  characterize all  boundary conditions in a typical  room. Further,  simulation time can be prohibitive if it is evaluated over a broadband spectrum.  Moreover, the whole simulation needs to be repeated for every new form factor of the device under test.  
To address the computational complexity, the image source method \cite{imagemethod} has been widely used to approximate point-to-point room acoustics. It utilizes the ray tracing concept \cite{RoomAcousticsBook} to significantly reduce  the modeling and computational complexity of brute-force simulation.  Though simple and effective in some scenarios, the image source method has few limitations. For example, it has poor approximation at low frequencies, and it cannot model small surfaces (as compared to wavelength), e.g., furniture, and rough surfaces, e.g., curtains. 

In this work, we describe a novel procedure that combines empirical and simulation methods to provide a balanced tradeoff between the two approaches for sound-field synthesis.  It splits the sound-field into two independent components: \emph{room} component, and \emph{device} component, such that the overall sound-field is the composite mapping of the two components.  The room component captures the room impact at an interior point, due to a predefined sound source in the room. This is represented   as a superposition of acoustic plane waves, which is computed using a single measurement with a large microphone array. The device component is computed using acoustic simulation or anechoic measurements to evaluate the \emph{fingerprint} of each acoustic plane wave on the device surface (as measured at the microphone array mounted on the device surface). The overall acoustic pressure on the device surface when placed at an interior point in a room is computed by plugging in the computed device fingerprints into the acoustic plane wave representation at that point. This arrangement provides an efficient representation of room acoustics that allows  reusing room information with devices of different form factor when tested in the same room. Therefore, it enables the concept of \emph{room database}, which contains abstract room acoustics information that is \emph{independent} of the device under test.   Likewise, it allows reusing the same device component with different rooms. 
To enable the proposed method, we develop a general procedure to compute the plane wave decomposition at a point in a room by applying sparse recovery techniques on an audio capture with a large microphone array. We also utilize the  \emph{device dictionary} concept, that captures the acoustic behavior of general microphone array mounted on a rigid surface of arbitrary form factor \cite{chhetri2019acoustic}.
The proposed methodology is rigorously validated across many rooms and many devices with different form factors and microphone array geometry. The synthetic RIR is shown to  match the true RIR, in the least square sense, over a broadband spectrum up to $8$ kHz. The synthesis methodology is also shown to closely resemble real measurements in evaluating higher level metrics, e.g., word error rate, and false rejection rate.

The acoustic plane wave expansion has been used in earlier work with model-based sound-field reconstruction, e.g., \cite{hahmann2021spatial, verburg2018reconstruction, koyama2018sparse, iijima2021binaural}, where acoustic plane waves are used as kernels for sound-field reconstruction. The plane wave expansion is interpolated with free-field propagation model to reproduce the sound field within a convex source-free zone. The plane wave expansion is computed from measurements  of an array of microphones placed at the zone perimeter. The computation of the expansion is done either through spherical harmonics or using sparse recovery techniques. In this work, we study a different problem of reproducing the sound field on a rigid surface that is placed at the same point in the room. A key contribution of the current work is utilizing the device dictionary concept for sound synthesis. This enables the generalization of the sound-field production to a rigid surface with an arbitrary form factor and microphone array size.

The paper is organized as follows. In section \ref{sec:foundation}, we lay down the theoretical foundation of the work. The details of the proposed framework are described in section \ref{sec:framework}. Then, we present the validation results in section \ref{sec:validation}. Finally, we describe few engineering applications in section \ref{sec:app}. The following notations are used throughout the paper. A bold lower-case letter denotes a column vector, while a bold upper-case letter denotes a matrix. 
$M$ always refers to the number of microphones.  The independent variables $t$ and $\omega$ refer to time and frequency respectively.
Additional notations are introduced when needed.

\section{\label{sec:foundation}Foundations}
\subsection{Acoustic Plane Waves}

Acoustic plane waves are eigenfunctions of the homogenous Helmholtz equation. Hence, they constitute a powerful tool for analyzing the wave equation. Further, a plane wave is a good approximation of the wave-field emanating from a far-field point source \cite{teutsch2007modal}. The acoustic pressure of a plane wave with vector wave number $\bf{k}(\theta, \phi)$ (where $\theta$ and $\phi$ correspond respectively to polar azimuth and elevation of the direction of propagation) is defined at a point  ${\bf{r}} = (x,y,z)$ in the three dimensional space  as \cite{FourierAcoustics}:
\begin{equation}
\psi({\bf{\omega, \theta, \phi, r}}) \triangleq p_0(\omega) e^{-j {\bf{k}}^T{\bf{r}}}   \label{eq:pw_def}
\end{equation} 
where $p_0(\omega)$ is a real-valued frequency dependent scaling.
The plane wave decomposition  has been used for approximating point-source seismic recording \cite{treitel1982plane, pwd2, zhou1994linear}, and sound field reproduction \cite{teutsch2007modal, kirkeby1993reproduction, ward2001reproduction, park2005sound}. 
A local solution to the homogenous Helmholtz equation can be approximated by a linear superposition of plane waves of different angles of the form \cite{pwd2, pwd1}:
\begin{equation}
p(\omega,{\bf{r}}) = \sum_{l \in \Lambda} \alpha_l(\omega) \ \psi\left(\omega, \theta_l, \phi_l, \bf{r}\right) \label{eq:gen_pw_sol}
\end{equation}
where $\Lambda$ is a set of indices that defines the directions of plane waves $\{\theta_l, \phi_l\}$,  each $\psi(.)$ is a plane wave as in \eqref{eq:pw_def}, and $\{\alpha_l\}$ are complex-valued scaling factors.  We will refer to the wave-field in \eqref{eq:gen_pw_sol} as the free-field acoustic pressure. The decision variables in this approximation are $\left\{\Lambda, \{\alpha_l\}_{l\in\Lambda}\right\}$. 
\subsection{Device Acoustic Dictionary}\label{sec:dict}
Generalizing the free-field plane wave expansion in \eqref{eq:gen_pw_sol} to include the scattering  due to the device surface, requires computing the device acoustic response to each plane wave. The device response to all plane waves in the three-dimensional space is collectively referred to as the device \emph{acoustic dictionary}.
The \emph{total} wave-field at any point on the device surface when the device is impinged by an incident plane wave $\psi(\omega, \theta, \phi, {\bf{r}})$ has the general form:
\begin{equation}
p_t(\omega, \theta, \phi, {\bf{r}}) = \psi(\omega, \theta, \phi, {\bf{r}}) + p_s(\omega, \theta, \phi, {\bf{r}})  \label{eq:pt_df}
\end{equation} 
where $p_t$ and $p_s$ refer to the total and scattered wave-field respectively. $p_t$ can be computed numerically by inserting \eqref{eq:pt_df} in the Helmholtz equation and solving for $p_s$ with appropriate boundary conditions. If a microphone array of size $M$ is mounted on the device surface, and the microphone port size is much smaller than the wavelength, then each microphone can be approximated by a point on the device surface. In this case, the total field, ${\mathbf{p}}_t(\omega, \theta, \phi)$, at the microphone array, due to an incident plane wave $\psi(\omega, \theta, \phi, {\bf{r}})$,  is a vector of size $M$ whose entries are the corresponding total field at the coordinate values, $\bf{r}$, of each individual microphone. The device acoustic dictionary of a device is composed of vectors of total acoustic wave-field. The device acoustic dictionary is computed using numerical acoustic simulation with Finite Element Method (FEM) or Boundary Element Method (BEM) with device CAD to specify the device surface. The details and validation results are described in \cite{chhetri2019acoustic}.

An entry of the device dictionary can be  either measured in anechoic room with single-frequency far-field sources, or computed numerically by solving the Helmholtz equation on the device surface  with background plane-wave using the device CAD model. Both methods yield same result, but the numerical method has much lower cost and it is less error-prone because it does not require human labor.

\begin{figure}[h] 
	\centering
	\includegraphics[draft=false, width=9cm, height=8.5cm, ,trim=30mm 10mm 30mm 15mm,clip]{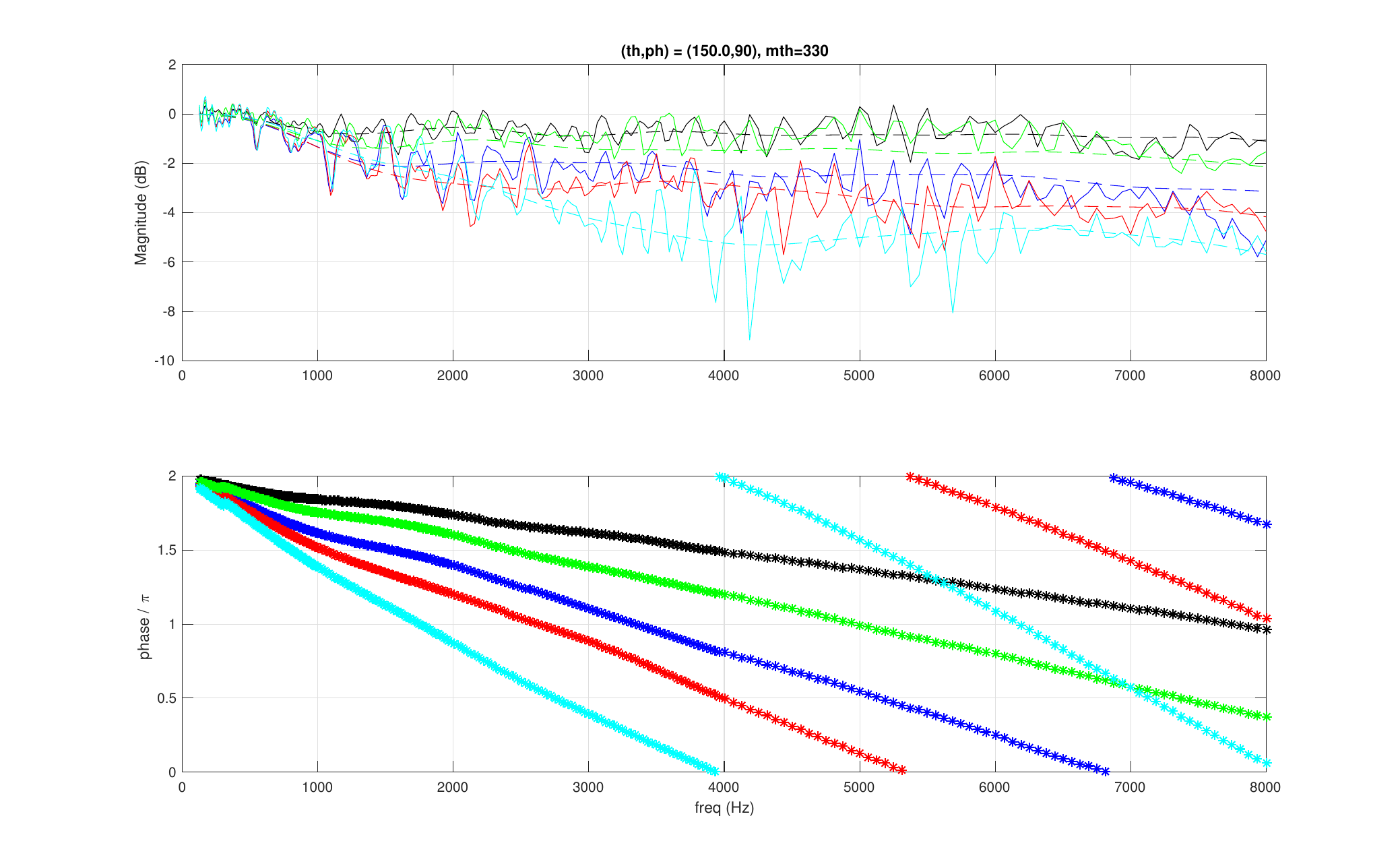}
	\caption{(top) Measured (solid line) and simulated (dotted line) total field of a  microphone array mounted on a sphere of PW, (bottom) phase error, with azimuth = $150^\circ$, elevation = $90^\circ$.}
	\label{fig:comsol_ver}
\end{figure}

For the numerical method, each entry in the device dictionary is computed by solving the Helmholtz equation, using Finite Element Method (FEM) or Boundary Element Method (BEM) techniques, for the total field at the microphones with the given background plane wave.  The device CAD is used to specify the surface, which is modeled as sound hard boundary. To have a true background plane-wave, the external boundary should be open and non-reflecting. In our model, the device is enclosed by a closed boundary, e.g., a cylinder or a spherical surface. To mimic open-ended boundary we use Perfectly Matched Layer (PML),  which defines a special absorbing domain that eliminates reflection and refractions in the internal domain that encloses the device \cite{berenger1994perfectly}. Standard packages for solving  partial differential equations, e.g.,  \cite{COMSOL} are used, and the simulation is rigorously validated  with measured acoustic pressure on different form-factors. 
In Fig. \ref{fig:comsol_ver}, we show an example of the frequency amplitude and phase of the inter-channel transfer function of both simulated and anechoic measured response for a microphone array mounted on a sphere. The reference channel in the inter-channel transfer function is the first microphone that is hit by the plane wave.  The phase plot at the bottom is the phase \emph{error} between the measured and simulated response. In the ideal case, the magnitude response of the measured and simulated transfer function should coincide, and the phase error is identically zero. The matching of the magnitude response is quite clear and ripples in the measure response is due to the impact of minor reflections in the anechoic room. Similarly, the phase error cannot be identically zero in practice because of the finite geometric precision in the position in the anechoic room, which  results in unavoidable \emph{linear} phase error that is shown in the phase plot. More validation examples were described in \cite{chhetri2019acoustic}. In  \cite{wiener1949diffraction,spence1948diffraction}, comparisons between simulated and theoretical acoustic pressure responses were presented.

The output of the above model is the plane wave dictionary of the device 
\begin{equation}
\mathcal{D} \triangleq \{\boldsymbol{\beta}_l(\omega) \triangleq {\bf{p}}_t(\omega, \theta_l, \phi_l): \forall \ \omega,l\}
\end{equation}
where each entry in the dictionary is a vector of length $M$, and each element in the vector is the total acoustic pressure at one microphone in the microphone array when a plane wave with ${\bf{k}}(\theta_l, \phi_l)$ hits the device. The dictionary also covers all frequencies of interest  typically up to $8$ kHz. Note that, the acoustic dictionary can accommodate any form factor of the surface, and any geometry of the microphone array as it utilizes acoustic simulation with the CAD of the device surface, and the coordinates of the microphone array.


\section{Proposed Framework}\label{sec:framework}
\subsection{Overview}\label{sec:overview}
The proposed sound synthesis methodology generalizes  the plane wave expansion in \eqref{eq:gen_pw_sol}, which summarizes  room acoustics at a point in a room, to include the impact of scattering due to the device surface. This generalization  yields the combined acoustic effect of the room and the scattering on the device surface. If  the device dimensions are much smaller than the room dimensions, then  \emph{secondary} reflections due to the device surface are negligible, and the device impact on the room acoustics could be ignored. Hence, even after introducing the device into the room, the directions and weights of the free-field acoustic plane waves in  \eqref{eq:gen_pw_sol} do not change. However, because of the device surface, each plane wave $\psi\left(\omega,\theta_l, \phi_l, {\bf{r}}\right)$ in  \eqref{eq:gen_pw_sol}, has an acoustic \emph{fingerprint} at the device microphone array, which is the corresponding entry in the device dictionary,  $\boldsymbol{\beta}_l(\omega)$. Hence, if the device is placed at a point in the room whose sound-field is expressed as in \eqref{eq:gen_pw_sol}, then the observed sound field vector at the device microphone array  is
\begin{equation}
{\mathbf{p}}(\omega) = \sum_{l\in \Lambda} \alpha_l \ \boldsymbol{\beta}_l(\omega) \label{eq:comb_pw_sol}
\end{equation}
The transition from the sound-field in \eqref{eq:gen_pw_sol} in the absence of the device to the sound field in \eqref{eq:comb_pw_sol} in the presence of the device is the technical foundation of the proposed synthesis framework. This transition is enabled by the linearity of the wave equation and the introduction of the device acoustic dictionary as described in section \ref{sec:dict}. 
 
Note that, if another device with acoustic dictionary $\mathcal{D}^{(2)} \triangleq \{\boldsymbol{\beta}_l^{(2)}(\omega): \forall \ \omega,l\}$ is placed at the same point in the room, then the observed sound-field at the second microphone array can be expressed as 
\begin{equation}
{\mathbf{p}}^{(2)}(\omega) = \sum_{l \in \Lambda} \alpha_l \ \boldsymbol{\beta}_l^{(2)}(\omega) 
\end{equation}
 where only the mapping through device dictionary changes, while the directions, $\Lambda$, and weights, $\{\alpha_l\}$, of constituent plane waves do not change. This is the essence of the proposed methodology that separates room and device components.
 Hence, the three steps for sound-field synthesis at a microphone array mounted on a device that is placed at a point in the room are as follows:
 \begin{enumerate}
 \item Compute the free-field plane wave expansion at a point as in \eqref{eq:gen_pw_sol}. This summarizes  room acoustics at a point in the room. A procedure for computing this expansion is described in section \ref{sec:pwd}. This process  is repeated for every new point in a room, and it is independent of the device under test.
 \item Generate the acoustic dictionary of the device under test as described in section \ref{sec:dict}. This is computed once per device and it is independent of the room.
 \item For each room position, combine the plane wave expansion with the device dictionary as in \eqref{eq:comb_pw_sol} to synthesize the sound-field at the device microphone array.
 \end{enumerate}
 Repeating step $1$  above  for multiple rooms and multiple positions within each room generates a \emph{room database}. This database is generated only once, then it could be reused in evaluating and generating data for new devices. 
\subsection{Acoustic Plane Wave Decomposition} \label{sec:pwd}
The main technical hurdle in the proposed framework is computing the plane wave expansion  \eqref{eq:gen_pw_sol} at a point in a room with a source signal emanating from another point in the room. In the proposed framework, this is computed through a data capture using a large microphone array of $32$ microphones mounted on a sphere (EigenMike) \cite{acoustics2013em32}. The large microphone array  is necessary to mitigate the creation of an underdetermined system of equations in recovering the constituent plane waves. It was found experimentally that $20$ to $30$ plane waves are sufficient for an accurate approximation of the sound field  (with reconstruction error less than $-30$ dB for frequencies up to $8$ kHz).
The plane wave decomposition problem is formulated as an optimization problem whose objective is minimizing the difference in the least square sense between observed and synthesized sound fields. If the observed sound field of the EigenMike at frequency $\omega$, is ${\mathbf{y}}(\omega)$, then the objective function has the form
\begin{equation}
J = \int_\omega \|{\mathbf{y}}(\omega) - \sum_{l \in \Lambda} \alpha_l(\omega)  \bar{\boldsymbol{\beta}}_l(\omega)\|^2 + \lambda \sum_{l \in \Lambda} |\alpha_l(\omega) |  \label{eq:opt}
\end{equation}
where $\left\{\bar{\boldsymbol{\beta}}_l(\omega)\right\}$ are the entries of the EigenMike acoustic dictionary at frequency $\omega$. The decision variables are the set of indices  $\Lambda$, and the corresponding weights $\{\alpha_l\}$. The L1-regularization in \eqref{eq:opt} is added to stimulate a sparse solution as $|\Lambda|< 30$ is much smaller than the dictionary size, which is in the order of $10^3$. The objective function can be put in matrix form as:
\begin{equation}
J = \int_\omega \|{\mathbf{y}}(\omega) - {\mathbf{A}}(\omega) \ . \ \boldsymbol{\alpha} \|_2^2 + \lambda \ |\boldsymbol{\alpha}|_1. \label{eq:matform}
\end{equation}
where ${\mathbf{A}}$ is a matrix whose columns are the individual entries of the acoustic dictionary at frequency $\omega$, i.e., $ \{\bar{\boldsymbol{\beta}}_l(\omega)\}$.
The above problem is a form of the well-known LASSO optimization \cite{tibshirani1996regression} that is encountered in numerous sparse recovery problems in statistics and signal processing. Many efficient solutions have been proposed for this problem under various conditions \cite{tropp2010computational, hastie2019statistical}. The big microphone array size in the EigenMike provides much flexibility in solving \eqref{eq:matform} because the observation size is bigger than the number of nonzero components in $ \boldsymbol{\alpha}$. Note that, the optimization problem in \eqref{eq:matform} is solved only once for a given source/receiver position in a room, and it is solved offline. Therefore, it does not have constraints on computational complexity, memory, or latency. In our analysis, the orthogonal matching pursuit algorithm \cite{tropp2007signal} was used to recover $\Lambda$ and $ \boldsymbol{\alpha}$, though other existing solutions to the sparse recovery problem can be used with this formation. This was generalized for smaller microphone arrays of arbitrary geometry in \cite{awd_icassp}.

For a given source signal, the above procedure is repeated at each frequency $\omega$, and at each time frame to generate a time-frequency map of the active set $\Lambda (t,\omega) $ and the corresponding weights $\boldsymbol{\alpha} (t,\omega)$. To synthesize the sound field for another device with acoustic dictionary $\{{\boldsymbol{\beta}}_l(\omega)\}$ at this particular source/receiver position and source signal, the synthesis formula \eqref{eq:comb_pw_sol} is applied at each time-frequency cell with the corresponding parameters $\Lambda (t,\omega) $ and  $\boldsymbol{\alpha} (t,\omega)$. Generating the sound-field for an arbitrary source signal requires the computation of the room impulse response, which is described in the following section.
\subsection{Room Impulse Response (RIR) Computation} \label{sec:rir}
RIR aims at modeling the acoustic channel between source and receiver as a linear time-invariant system. The RIR combines both room acoustics and scattering due to device surface, and it is computed once for a given device and a given source/receiver positions in a room. It is a multichannel transfer function where the number of channels equals the size of the microphone array. For RIR computation, a special source signal that covers the whole frequency spectrum, e.g., white noise or Golay sequence \cite{foster1986impulse}, is utilized. 
For a source signal $x(t,\omega)$, the EigenMike is utilized to generate the time-frequency map of the plane wave decomposition as described in the previous section. For a device under test, this time-frequency map is combined with the device dictionary to generate the multichannel output signal ${\mathbf{y}}(t,\omega)$ as in \eqref{eq:comb_pw_sol}. The transfer function between $x(t,\omega)$ and ${\mathbf{y}}(t,\omega)$ is computed using system identification techniques. For example, by applying Wiener-Hopf equation in the frequency domain \cite{stoica1997introduction}, we get
\begin{equation}
\hat{\mathbf{h}}(\omega) = \frac{\mathbf{S}_{xy}(\omega)}{S_{xx}(\omega)}
\end{equation}
where $\hat{\mathbf{h}}(\omega) $ is the multichannel acoustic transfer function in the frequency domain, and 
\begin{eqnarray}
S_{xx}(\omega) &=& \mathbb{E}\left\{x^*(t,\omega)\ x(t,\omega)\right\} \\
{\mathbf{S}}_{xy}(\omega) &=& \mathbb{E}\left\{x^*(t,\omega)\ {\mathbf{y}}(t,\omega)\right\}
\end{eqnarray}
After RIR estimation for a device at a point in a room, the sound field for an arbitrary source signal $u(t,\omega)$ is computed as
\begin{equation}
\hat{\mathbf{y}}(t,\omega) = \hat{\mathbf{h}}(\omega)\ . \ u(t,\omega) \label{eq:synth-rir}
\end{equation}
Note that, the RIR computes only the part of the sound field that is correlated with the source signal, and it disregards the background ambient noise and other interferences in the room. 
To add background and/or diffuse noise to the synthesized output, a separate time-frequency map, ${\mathbf{b}}(t,\omega)$, is computed once as described in section \ref{sec:overview} with only background noise, then the synthesized sound-field in \eqref{eq:synth-rir} is modified to
\begin{equation}
\hat{\mathbf{y}}(t,\omega) = \hat{\mathbf{h}}(\omega)\ . \ u(t,\omega) + {\mathbf{b}}(t,\omega)
\end{equation}


\subsection{\label{sec:discussion}Discussion}

The proposed method is a combination of measurements (for room component) and simulation (for device component). 
A single measurement with a large microphone array is required per room position, and this measurement is reused for all devices. The measurement is  processed by plane wave decomposition to compute the time-frequency map of the acoustic decomposition that is combined with device dictionary to generate the total sound field.
Similarly, the device dictionary is computed once, and it is combined with any room to generate the total sound-field. The computational complexity for computing the broadband device dictionary is small because it is computed in an anechoic setup. Further, it is highly parallelizable because the same process is repeated at different frequencies and at different directions for plane wave.
The concept of splitting  room acoustics and device acoustics significantly reduces the measurement/simulation overhead. In addition to simplifying both room and device modeling,  abstracting room acoustics in a single measurement and device acoustics with the device dictionary enables reuse of either components with the other side.

The proposed approach provides an accurate approximation of room acoustics and alleviates the need for full room  simulation whose complexity is prohibitive at high frequencies  for a regular-size room. Further, this single room measurement eliminates the need to model the room interior surfaces, which can also be an overly time-consuming process.  As compared to the image source method, the proposed method addresses all the limitations outlined in section \ref{sec:introduction} as follows:
 \begin{enumerate}
 \item The plane-wave expansion model in \eqref{eq:comb_pw_sol} is valid at all frequencies. 
 \item The impact of device scattering is incorporated through the device dictionary in  sound-field synthesis.
 \item The impact of all boundary conditions in the room is inherently included in the plane-wave expansion in \eqref{eq:gen_pw_sol}. It automatically accounts for all surfaces in the room without explicitly modeling them.
 \end{enumerate}
Note that, it is possible to combine the device acoustic dictionary  with the image source method  \cite{imagemethod} to further educe  complexity at the cost of lower accuracy. In the image source method, the concept of a sound wave is replaced by sound rays; which is a small portion of a spherical wave with vanishing aperture \cite{RoomAcousticsBook}. These sound rays from the image source method can be regarded as a crude approximation of acoustic plane waves in \eqref{eq:gen_pw_sol}; which eliminates the need for room measurements. If these sound rays are combined with device dictionary, then it extends the image source method to account for scattering due to the device surface. However, it still inherits the other gaps of the image source method as previously outlined  in section \ref{sec:introduction}.
  
\section{Experimental Validation}\label{sec:validation}
\begin{figure}[h] 
	\centering
	\includegraphics[width=8.0cm, height=5.5cm, ,trim=0mm 2mm 0mm 0mm,clip]{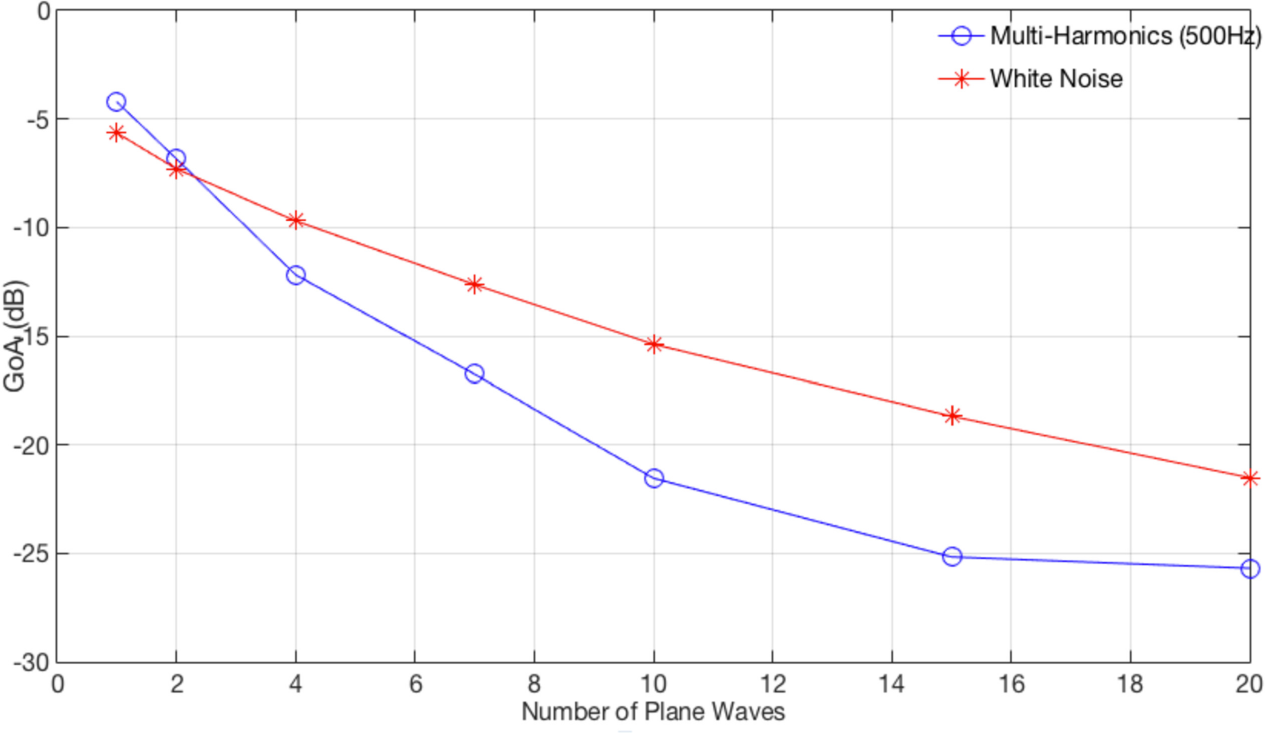}
	\caption{EigenMike reconstruction error versus the number of plane waves in the plane wave decomposition}
	\label{fig:eigmike_rec}
\end{figure}

The first experiment aimed at validating the plane wave decomposition procedure as described in section \ref{sec:pwd}. In Fig. \ref{fig:eigmike_rec}, we showed the reconstruction error of the EigenMike for two different source signals versus the number of plane wave in the expansion. The Goodness of Approximation, GoA, (or reconstruction error) is defined as:
\begin{equation}
\text{GoA} \triangleq  \frac{\int_\omega \|{\mathbf{y}}(\omega) - \sum_{l \in \Lambda} \alpha_l(\omega)  \bar{\boldsymbol{\beta}}_l(\omega)\|^2 }{\int_\omega \|{\mathbf{y}}(\omega)\|^2}
\end{equation}
where ${\mathbf{y}}(\omega)$ is the observed sound-field at the EigenMike. This was evaluated over a frequency range up to 8 kHz. As noted from the figure, a small number of plane waves is sufficient for sound field approximation with error less than $-20$ dB.

\begin{figure}[h] 
	\centering
	\includegraphics[width=11cm, height=9.5cm, ,trim=20mm 8mm 20mm 8mm,clip]{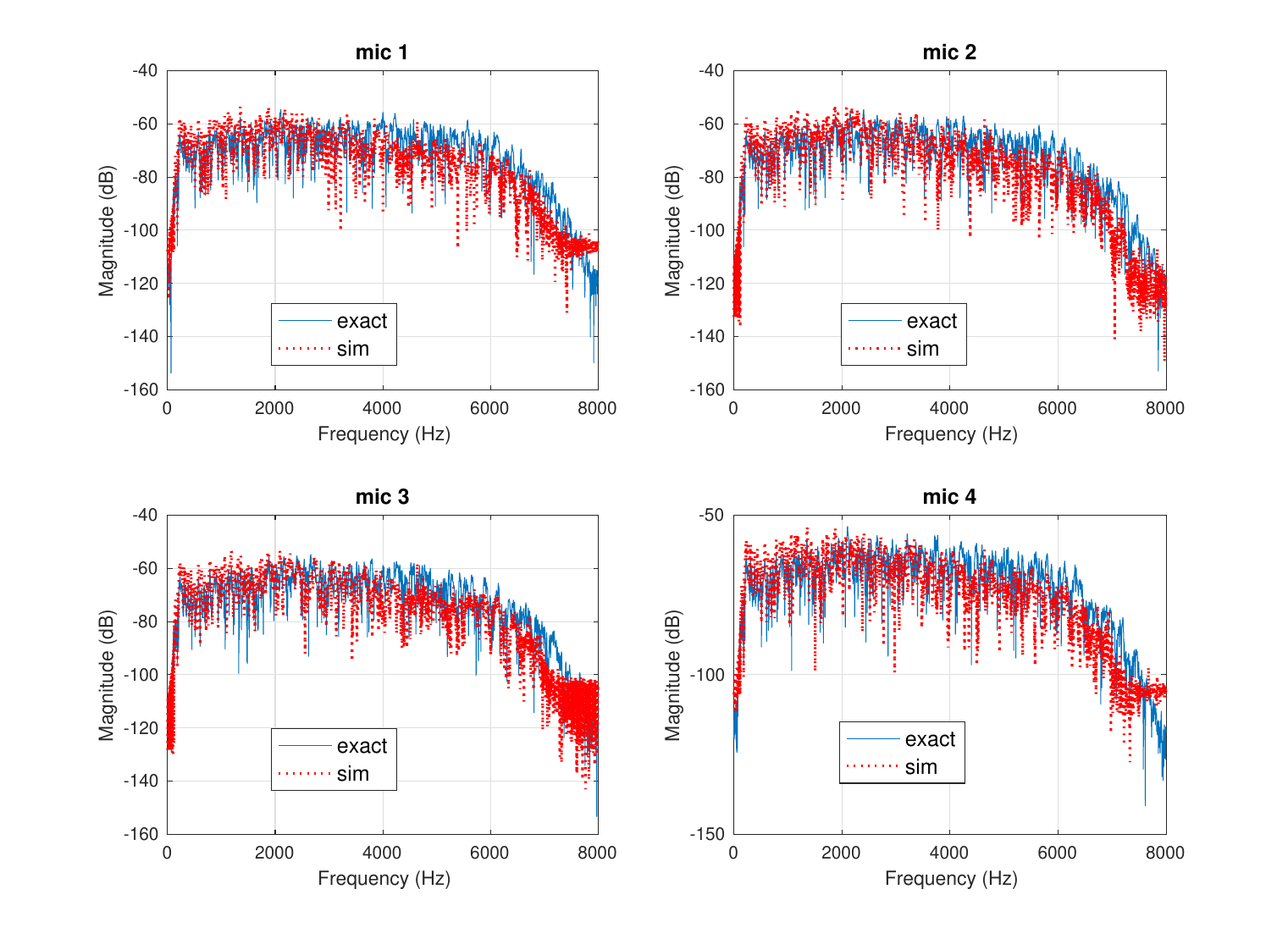}
	\caption{An example of frequency response of measured and synthetic RIR for a microphone array of size 4 on a slated sphere surface}
	\label{fig:fd_comparison}
\end{figure}

\begin{figure}[h] 
	\centering
	\includegraphics[width=11cm, height=9.5cm, ,trim=20mm 8mm 20mm 8mm,clip]{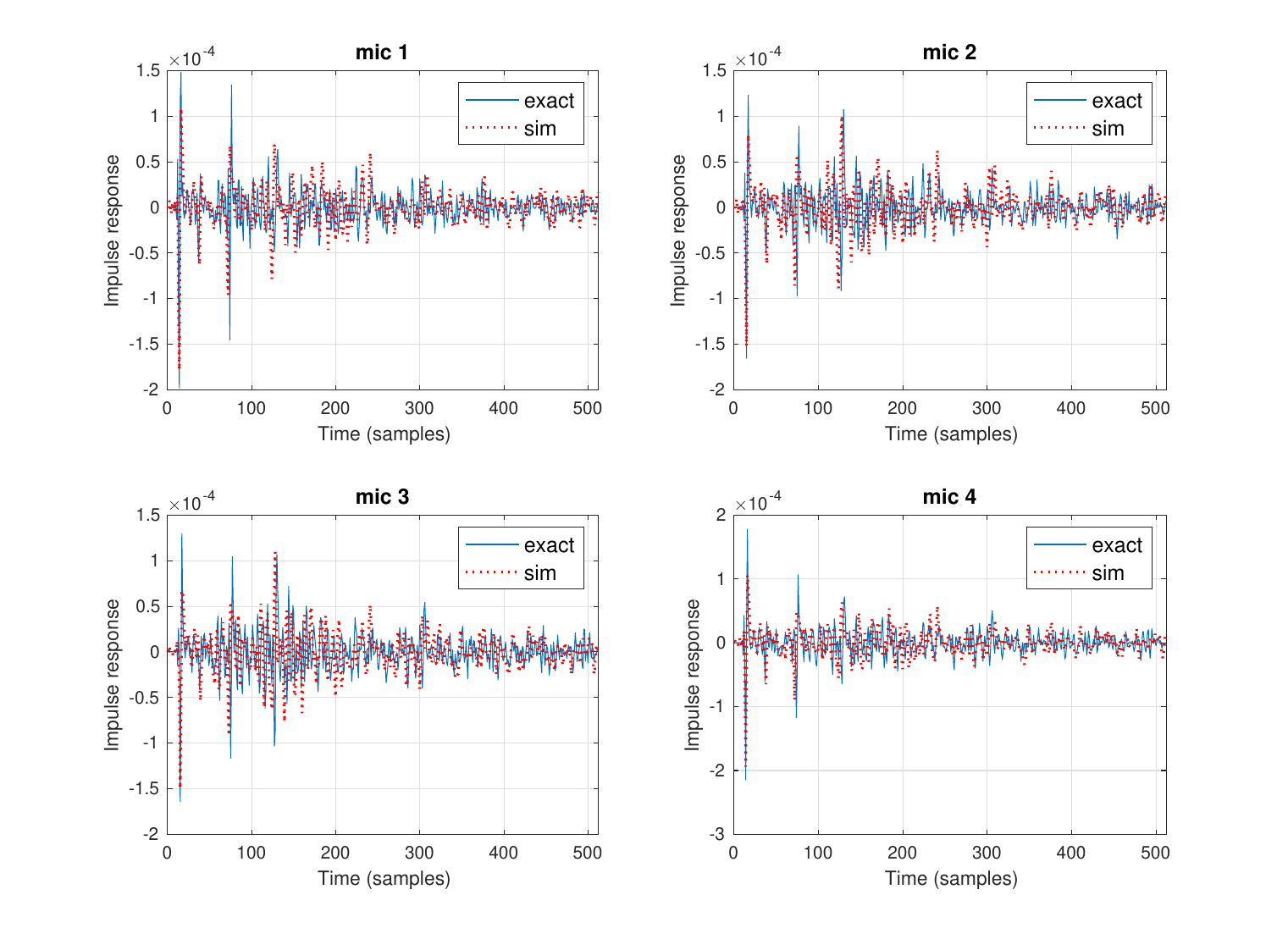}
	\caption{An example of impulse response of measured and synthetic for microphone array of size 4  on a slated sphere surface}
	\label{fig:td_comparison}
\end{figure}

In the following set of experiments, the RIR procedure as described in section \ref{sec:rir} was evaluated. The experiments were conducted in three different rooms with furniture that resemble typical bedrooms and living rooms, and in $24$ different positions within the three rooms. The EigenMike was placed in all positions to compute the room component  that is combined with device dictionary to generate the synthetic RIR. Four other devices with different form factors and microphone array geometries were placed later at the same positions to compute the true RIR. Two devices had cylindrical form factor, one had cube-like shape, and the fourth was a slated sphere. The $24$ test positions covered different room positions: middle of the room, next to a wall, and at a corner. For each position, the origins of the EigenMike and other devices were aligned precisely using laser beams. The measured and synthetic RIR are computed as described in section \ref{sec:rir}. In all cases, there existed strong resemblance between measured and synthetic RIR at all frequencies,  and the reconstruction signal-to-noise ratio (SNR) is between $19$ and $23$ dB. An example of the transfer function and the impulse response of the measured and synthetic RIR is shown respectively in Figures \ref{fig:fd_comparison} and \ref{fig:td_comparison}. This is a typical behavior at all positions and devices.

In the third set of experiments, we studied the impact of  mismatch between true and synthetic RIR in evaluating high level data metrics. For this test, we evaluated the False Rejection Rate (FRR) of a keyword in the source signal. The true FRR was computed by processing the device signal after the convolution of the source signal and the true RIR. Similarly, the synthetic FRR was computed from the convolution of the same source signal and synthetic RIR. The relative absolute FRR is defined as
\begin{equation}
\text{Relative FRR  Error} \triangleq \frac{|\ \text{FRR}_{true} - \text{FRR}_{synthetic}\ |}{\text{FRR}_{true}}
\end{equation}

\begin{figure}[h] 
	\centering
	\includegraphics[width=9.0cm, height=6.0cm, ,trim=15mm 0mm 15mm 7mm,clip]{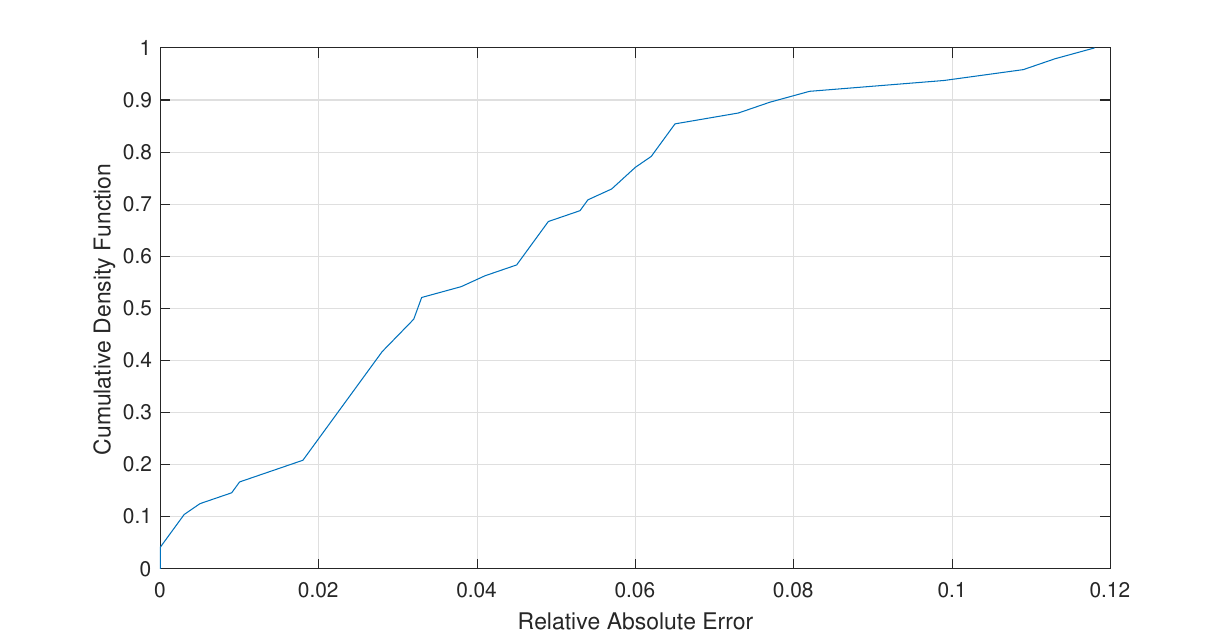}
	\caption{Cumulative density function (CDF) of the relative absolute FRR error}
	\label{fig:cdf}
\end{figure}

The cumulative density function of the relative absolute error (of all $24$ room positions and all devices) is shown in Fig. \ref{fig:cdf}. The $95$-percentile relative error between true and synthetic FRR is less than $10\%$.

\section{\label{sec:app}Relevant Applications}
\subsection{Data Generation}
The primary application of the proposed sound synthesis procedure is generating synthetic data to compute performance analytics for the device under test, e.g., False Rejection Rate (FRR), Word Error Rate (WER), and audio quality metrics. It provides a low-cost high-quality alternative to real data collection; which is  an expensive and time-consuming process. Further, the proposed framework provides control of the environment conditions for customized usage scenarios. The synthetic data can also be used to augment real data collection for training deep learning acoustic models, which require large amount of data under diverse acoustic conditions. 
Note that, the proposed methodology requires only the device CAD (to compute the device dictionary), and does not need  the actual device hardware. Therefore, it could be integrated with the hardware design process to evaluate different metrics at early design stages without building hardware prototypes.

%

\subsection{Microphone Array Processing}
The traditional model for microphone array processing in the literature utilizes free-field steering vectors that capture only  phase differences due to wave propagation and neglect the magnitude component due to scattering with device surface \cite{MA_book, benesty2008microphone}. This limitation results in undesirable effects with beamforming, e.g., spatial aliasing. 
Rather, the acoustic dictionary, as described in section \ref{sec:dict},  is a generalization of free-field steering vectors that captures device response to acoustic plane wave. 
The entries of the acoustic dictionary provide more accurate steering vectors because they have both magnitude and phase components.  Therefore, beamformer design with steering vectors from acoustic dictionary provides significantly better directivity in real room setting \cite{chhetri2019acoustic}.
Further, the plane wave decomposition in \eqref{eq:comb_pw_sol}  incorporates scattering at the device surface; hence, it provides an accurate acoustic map of the device surrounding. It can be regarded as an invertible spatial transformation that maps microphone array observations into  spatial components. This enables many applications related to microphone array processing, e.g., sound source localization, sound source separation, and dereverberation.

\section{Conclusions}
The proposed framework enables generating accurate multichannel audio data for a general microphone array mounted on a rigid surface of an arbitrary form factor using only the CAD model of the surface and the coordinates of the microphone array. 
It is a balanced approach between measurement-based methods and simulation-based methods.  
The framework significantly reduces the cost of hardware validation and data collection. Its effectiveness is established by rigorous validation with physical data under diverse scenarios.

The key contribution of the proposed method is the decoupling of the room impact and the device-dependent impact in estimating the sound-field at the microphone array. This decoupling enables reusing the room  component with different hardware designs, as if the corresponding devices are placed in the same room location. The room acoustics modeling is performed using plane wave decomposition, after collecting room measurements with a large microphone array to achieve high accuracy modeling as shown in the experimental validation.  Device modeling through the device acoustic  dictionary provides a general model that works with any microphone array geometry and any form factor of the  surface.  Though the discussion focused primarily on far-field sources and acoustic plane waves, it can be straightforwardly extended to the near field case by appending the device dictionary with acoustic spherical waves.

The contributions of this work are summarized as follows:
\begin{enumerate}
\item A methodology for decoupling and combining room acoustics and impact of device surface for accurate sound field realization.
\item A novel algorithm to compute plane wave decomposition of a point in a room based on measurement with a large microphone array with sparse recovery techniques.
\item A methodology for characterizing device acoustics based on response to acoustic plane waves.
\item A framework for data generation based on room impulse response that is derived from the proposed sound synthesis procedure.
\end{enumerate}
We also highlighted few relevant applications with high impact that are based on the proposed framework; which are investigated in details  in future publications.








\bibliographystyle{IEEEtran}
\bibliography{refs}

\begin{thebibliography}{10}
\providecommand{\url}[1]{#1}
\csname url@samestyle\endcsname
\providecommand{\newblock}{\relax}
\providecommand{\bibinfo}[2]{#2}
\providecommand{\BIBentrySTDinterwordspacing}{\spaceskip=0pt\relax}
\providecommand{\BIBentryALTinterwordstretchfactor}{4}
\providecommand{\BIBentryALTinterwordspacing}{\spaceskip=\fontdimen2\font plus
\BIBentryALTinterwordstretchfactor\fontdimen3\font minus
  \fontdimen4\font\relax}
\providecommand{\BIBforeignlanguage}[2]{{%
\expandafter\ifx\csname l@#1\endcsname\relax
\typeout{** WARNING: IEEEtran.bst: No hyphenation pattern has been}%
\typeout{** loaded for the language `#1'. Using the pattern for}%
\typeout{** the default language instead.}%
\else
\language=\csname l@#1\endcsname
\fi
#2}}
\providecommand{\BIBdecl}{\relax}
\BIBdecl

\bibitem{RoomAcousticsBook}
H.~Kuttruff, \emph{Room acoustics}, 4th~ed.\hskip 1em plus 0.5em minus
  0.4em\relax {CRC} Press, 2000.

\bibitem{imagemethod}
J.~B. Allen and D.~A. Berkley, ``Image method for efficiently simulating
  small-room acoustics,'' \emph{The Journal of the Acoustical Society of
  America}, vol.~65, no.~4, pp. 943--950, 1979.

\bibitem{chhetri2019acoustic}
A.~Chhetri, M.~Mansour, W.~Kim, and G.~Pan, ``On acoustic modeling for
  broadband beamforming,'' in \emph{2019 27th European Signal Processing
  Conference (EUSIPCO)}.\hskip 1em plus 0.5em minus 0.4em\relax IEEE, 2019, pp.
  1--5.

\bibitem{hahmann2021spatial}
M.~Hahmann, S.~A. Verburg, and E.~Fernandez-Grande, ``Spatial reconstruction of
  sound fields using local and data-driven functions,'' \emph{The Journal of
  the Acoustical Society of America}, vol. 150, no.~6, pp. 4417--4428, 2021.

\bibitem{verburg2018reconstruction}
S.~A. Verburg and E.~Fernandez-Grande, ``Reconstruction of the sound field in a
  room using compressive sensing,'' \emph{The Journal of the Acoustical Society
  of America}, vol. 143, no.~6, pp. 3770--3779, 2018.

\bibitem{koyama2018sparse}
S.~Koyama, N.~Murata, and H.~Saruwatari, ``Sparse sound field decomposition for
  super-resolution in recording and reproduction,'' \emph{The Journal of the
  Acoustical Society of America}, vol. 143, no.~6, pp. 3780--3795, 2018.

\bibitem{iijima2021binaural}
N.~Iijima, S.~Koyama, and H.~Saruwatari, ``Binaural rendering from microphone
  array signals of arbitrary geometry,'' \emph{The Journal of the Acoustical
  Society of America}, vol. 150, no.~4, pp. 2479--2491, 2021.

\bibitem{teutsch2007modal}
H.~Teutsch, \emph{Modal array signal processing: principles and applications of
  acoustic wavefield decomposition}.\hskip 1em plus 0.5em minus 0.4em\relax
  Springer, 2007, vol. 348.

\bibitem{FourierAcoustics}
E.~G. Williams, \emph{Fourier acoustics: sound radiation and nearfield
  acoustical holography}.\hskip 1em plus 0.5em minus 0.4em\relax Academic
  press, 1999.

\bibitem{treitel1982plane}
S.~Treitel, P.~Gutowski, and D.~Wagner, ``Plane-wave decomposition of
  seismograms,'' \emph{Geophysics}, vol.~47, no.~10, pp. 1375--1401, 1982.

\bibitem{pwd2}
O.~Yilmaz and M.~T. Taner, ``Discrete plane-wave decomposition by
  least-mean-square-error method,'' \emph{Geophysics}, vol.~59, no.~6, pp.
  973--982, 1994.

\bibitem{zhou1994linear}
B.~Zhou and S.~A. Greenhalgh, ``Linear and parabolic $\tau$-p transforms
  revisited,'' \emph{Geophysics}, vol.~59, no.~7, pp. 1133--1149, 1994.

\bibitem{kirkeby1993reproduction}
O.~Kirkeby and P.~A. Nelson, ``Reproduction of plane wave sound fields,''
  \emph{The Journal of the Acoustical Society of America}, vol.~94, no.~5, pp.
  2992--3000, 1993.

\bibitem{ward2001reproduction}
D.~B. Ward and T.~D. Abhayapala, ``Reproduction of a plane-wave sound field
  using an array of loudspeakers,'' \emph{IEEE Transactions on speech and audio
  processing}, vol.~9, no.~6, pp. 697--707, 2001.

\bibitem{park2005sound}
M.~Park and B.~Rafaely, ``Sound-field analysis by plane-wave decomposition
  using spherical microphone array,'' \emph{The Journal of the Acoustical
  Society of America}, vol. 118, no.~5, pp. 3094--3103, 2005.

\bibitem{pwd1}
A.~Moiola, R.~Hiptmair, and I.~Perugia, ``Plane wave approximation of
  homogeneous helmholtz solutions,'' \emph{Zeitschrift f{\"u}r angewandte
  Mathematik und Physik}, vol.~62, no.~5, p. 809, 2011.

\bibitem{berenger1994perfectly}
J.-P. Berenger, ``A perfectly matched layer for the absorption of
  electromagnetic waves,'' \emph{Journal of computational physics}, vol. 114,
  no.~2, pp. 185--200, 1994.

\bibitem{COMSOL}
{COMSOL Multiphysics}, ``Acoustic module--user guide,'' 2017.

\bibitem{wiener1949diffraction}
F.~M. Wiener, ``The diffraction of sound by rigid disks and rigid square
  plates,'' \emph{The Journal of the Acoustical Society of America}, vol.~21,
  no.~4, pp. 334--347, 1949.

\bibitem{spence1948diffraction}
R.~Spence, ``The diffraction of sound by circular disks and apertures,''
  \emph{The Journal of the Acoustical Society of America}, vol.~20, no.~4, pp.
  380--386, 1948.

\bibitem{acoustics2013em32}
Eigenmike, ``Em32 eigenmike microphone array release notes (v17. 0),'' \emph{MH
  Acoustics, USA}, 2013.

\bibitem{tibshirani1996regression}
R.~Tibshirani, ``Regression shrinkage and selection via the lasso,''
  \emph{Journal of the Royal Statistical Society: Series B (Methodological)},
  vol.~58, no.~1, pp. 267--288, 1996.

\bibitem{tropp2010computational}
J.~A. Tropp and S.~J. Wright, ``Computational methods for sparse solution of
  linear inverse problems,'' \emph{Proceedings of the IEEE}, vol.~98, no.~6,
  pp. 948--958, 2010.

\bibitem{hastie2019statistical}
T.~Hastie, R.~Tibshirani, and M.~Wainwright, \emph{Statistical learning with
  sparsity: the lasso and generalizations}.\hskip 1em plus 0.5em minus
  0.4em\relax Chapman and Hall/CRC, 2019.

\bibitem{tropp2007signal}
J.~A. Tropp and A.~C. Gilbert, ``Signal recovery from random measurements via
  orthogonal matching pursuit,'' \emph{IEEE Transactions on information
  theory}, vol.~53, no.~12, pp. 4655--4666, 2007.

\bibitem{awd_icassp}
M.~Mansour, ``Sparse recovery of acoustic waves,'' in \emph{ICASSP 2022-2022
  IEEE International Conference on Acoustics, Speech and Signal Processing
  (ICASSP)}.\hskip 1em plus 0.5em minus 0.4em\relax IEEE, 2022, pp. 5418--5422.

\bibitem{foster1986impulse}
S.~Foster, ``Impulse response measurement using golay codes,'' in
  \emph{Acoustics, Speech, and Signal Processing, IEEE International Conference
  on ICASSP'86.}, vol.~11.\hskip 1em plus 0.5em minus 0.4em\relax IEEE, 1986,
  pp. 929--932.

\bibitem{stoica1997introduction}
P.~Stoica and R.~L. Moses, \emph{Introduction to spectral analysis}.\hskip 1em
  plus 0.5em minus 0.4em\relax Pearson Education, 1997.

\bibitem{MA_book}
M.~Brandstein and D.~Ward, \emph{Microphone arrays: signal processing
  techniques and applications}.\hskip 1em plus 0.5em minus 0.4em\relax Springer
  Science \& Business Media, 2013.

\bibitem{benesty2008microphone}
J.~Benesty, J.~Chen, and Y.~Huang, \emph{Microphone array signal
  processing}.\hskip 1em plus 0.5em minus 0.4em\relax Springer Science \&
  Business Media, 2008, vol.~1.

\end{thebibliography}

\end{document}